# Outdoor Characterization of Solar Cells with Micro-structured Anti-Reflective Coating in a Concentrator Photovoltaic Module

Arnaud J. K. Leoga[1,2], Arnaud Ritou[1,2,3], Mathieu Blanchard[1,2], Lysandre Dirand[1,2], Yanis Prunier[1,2], Philippe St-Pierre[1,2], David Chuet[1,2], Philippe-Olivier Provost[1,2], Maïté Volatier[1,2], Vincent Aimez[1,2], Gwenaëlle Hamon[1,2], Abdelatif Jaouad[1,2], Christian Dubuc[3], Maxime Darnon[1,2]

*Abstract*— Micro-structured anti reflective coatings (ARC) have been identified as a promising solution to reduce optical losses in Concentrator Photovoltaics modules (CPV). We fabricated and tested in field a CPV modules made of 4 sub-modules with a concentration factor of 250×, that embed either solar cells with micro-structured encapsulating ARC or solar cells with multilayer ARC as a reference. The micro-structured encapsulating ARC was made of semi-buried silica beads in polydimethylsiloxane (PDMS). The module was in operation for 1 year in the severe climatic conditions of Sherbrooke, Quebec, Canada, before extracting the sub-modules performance under Concentrator Standard Operating Condition (CSOC).

An acceptance angle of ±0.78° was determined for all sub-modules, demonstrating that improving angular collection at the cell level has no significant impact on the angle of acceptance at the module level. We report an increase of 12 to 14% of the short-circuit current and of 15 to 19% of maximum power at CSOC for solar cells with a micro structured encapsulating ARC compared to the reference. Despite a sub-optimal module design, we report a sub-module efficiency of 29.7% at CSOC for a cell with micro-structured encapsulating ARC. This proves the potential of micro-structured encapsulating ARC to improve CPV system performance and shows promise of reliability for sumi-buried microbeads in PDMS as encapsulating ARC.

## I. Introduction

Concentrator photovoltaics (CPV) is the photovoltaic technology with the highest conversion efficiency at the system level. It relies on multi-junction solar cells that were originally developed for space application. To mitigate the high cost of multi-junction solar cells, concentrator photovoltaics uses optical elements (mirrors or lenses) to collect the sunlight and concentrate it onto a smaller area solar cell [1]. Current commercial CPV technologies reach more than 30% conversion efficiency at the system level, thanks to few mm$^2$ solar cells and Fresnel lenses with an optical concentration of typically 500×. Because of the presence of multiple optical interfaces combined with tracking error, optical efficiency losses due to reflection cannot be neglected and can be larger than 15% [2]. To reduce reflective losses at the surface of the triple junction solar cells, interferometric multi-layer anti-reflective coatings (ARC) such as $TiO_2/AlO_x$ [3] or $SiN/SiO_2$ [4] are deposited on the surface of the solar cells, with controlled thickness and optical index. These layers can eventually play the role of encapsulating layers to prevent interaction of the cell with moisture and improve the system reliability. However, such interferometric approach is designed for normal incidence, and is less effective when sunlight reaches the cell surface with an angular distribution, as it is the case in a CPV system. An alternative approach consists in micro/nano structuring the antireflective coating to provide a graded optical index on the cell [5,6]. Using modeling and indoor characterizations, we have shown that the short circuit current produced by an InGaP/InGaAs/Ge solar cell can be improved by 3.7 ±1%, by using 1 μm-diameter silica beads embedded into a Poly Dimethyl Siloxane (PDMS) layer—so-called micro-beads ARC [7]. In comparison with a $AlO_x/TiO_x/SiO_2$ encapsulating antireflective coating, indoor external quantum efficiency measurements of solar cells with micro-beads have demonstrated a 2.6% improvement in the triple junction solar cell short circuit current. We propose here to evaluate micro-beads ARC in real operating conditions. A dedicated CPV module with four 250× spherical lenses has been built [8]. Two solar cells with micro-beads ARC and two reference

[1]Institut Interdisciplinaire d'Innovation Technologique (3IT), Université de Sherbrooke, 3000 Boulevard Université, Sherbrooke, J1K 0A5 Québec, Canada

[2]Laboratoire Nanotechnologies Nanosystèmes (LN2) - CNRS IRL-3463, Institut Interdisciplinaire d'Innovation Technologique (3IT), Université de Sherbrooke, 3000 Boulevard Université, Sherbrooke, J1K 0A5 Québec, Canada

[3]Saint-Augustin Canada Electric Inc., Innovation and development of solar product, 75 rue d'Anvers, Saint-Augustin, Quebec Canada, G3A 1S5

Corresponding author: maxime.darnon@usherbrooke.ca

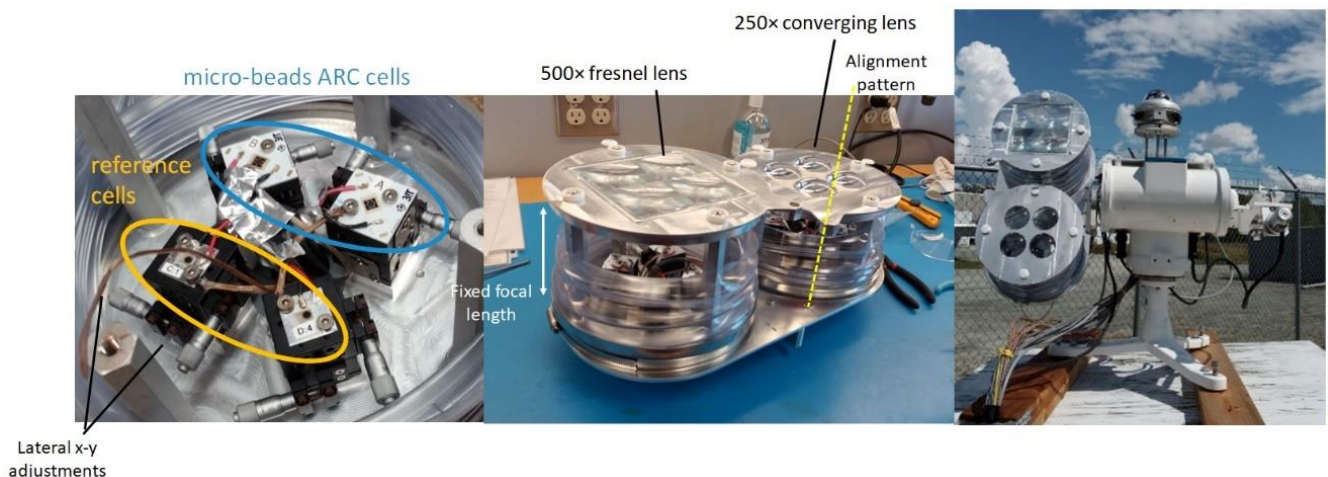

Figure 1: Pictures of (left) the inside of the test module, (middle) the complete test module with its Fresnel lens twin and (right) the module installed on the EKO tracker. Note that the Fresnel lens twin is not considered here.

solar cells with $AlO_x/TiO_x/SiO_2$ encapsulating ARC are integrated into the module. Each cell is independently characterized. The module has been installed for one year on an experimental sun tracker and exposed to harsh operating conditions in the solar park of Université de Sherbrooke, Quebec, Canada (including -30°C nights during winter and +30°C during summer).

## II. EXPERIMENTAL SECTIONS

### A. Solar cells

The solar cells are commercial triple junction solar cells from AZUR SPACE with an efficiency rated at 37.8% at 1000× (AM1.5D – 1000W/m²). The top electrode design has been optimized for operation under 500× concentration. The solar cells are hexagonal with an active area (excluding busbars) of 6.55 mm². The cells are coated with a bilayer $AlO_x/TiO_x$ ARC which thicknesses are optimized for minimizing reflection when used with an $SiO_2$ encapsulating layer. Two kinds of encapsulating layers are considered. As a reference, a 100 ±30 nm-thick layer of $SiO_2$ is deposited by atmospheric plasma enhanced chemical vapor deposition. The reference cells are bonded with a conductive epoxy onto an aluminum plate, and the front electrical contact is connected by gold wirebonding. The second kind of encapsulating layer (micro-beads) is made of 1 µm-diameter silica beads semi-buried into a 6 ±0.2 µm-thick PDMS layer (Sylgard 184). These cells with micro-beads are mounted on an Aluminum core printed circuit board (PCB) and the front contact is connected by gold wirebonding.

### B. Test module

The test module is made of two thick aluminum plates—one to support the lenses and one to support the cells—separated by aluminum posts. The lenses are plano-convex BK7 lenses from Thorlabs (LA1145) [9] glued using PDMS. These lenses have a focal distance of 74.8 mm and a diameter of 2 inches. The clear aperture of the lens is 16.42 cm². The lens is not coated by an ARC. Considering the active area of the solar cells, the optical concentration of the module is 250×. The cells on receiver are fixed onto micrometric plates for finetuning the cell-to-lens alignment. Each cell is individually connected to the electrical characterization setup. The module is made of 4 lens-cell couples, which will be referred to as sub-modules in the following. Two sub-modules (A and B) embedded solar cells with micro-beads while two other sub-modules (C and D) embedded reference cells. Once the test module is mounted onto the tracker, iterative corrections of the cell's positions combined with short circuit current measurements are performed to maximize the short circuit current of each sub-module, and therefore optimize the cell-to-lens alignment. Figure 1 shows pictures of the 250× module. We can see on Figure 1 that the module is twinned with a second module with Fresnel lenses. However, this second module is not considered here.

The test module is installed onto a STR-22G tracker from EKO (Figure 1) [10]. This tracker has a tracking accuracy of ±0.01°. For the first year of operation, the tracker was operating in astronomical mode (open-loop tracking). However, this tracking mode induces tracking errors that depend on the hour and day of operation. To minimize the tracking errors, we implemented a new tracking algorithm called MPPT (Maximum Power Point Tracking). It consists in measuring the short circuit current in 16 positions (4 x 4 resolution) of 2.25° range in azimuth and elevation centered on the astronomical position of the sun. From these measurements, we interpolate up to 100 positions (10 x 10 resolution)—this short circuit current map yielding the maximum short circuit current and the associated best position. While the sky is cloudless for a long enough period of time, the tracker then moves to this exact position before data acquisition. This tracking procedure is repeated for each sub-module before each measurement. Since the tracking procedure takes around 10s per sub-module, we consider the offset between the astronomical sun position and its real position to be constant over the 10s alignment and measurement time.

Based on the cell alignment procedure and the MPPT tracking algorithm, we can assume a perfect alignment of the cell, the lens, and the sun.

### C. Instrumentation

The electrical characterization of the sub-modules is performed using a 2601B SMU and a 3706A multiplexer from Keithley. Between each measurement, the solar cells are maintained in open circuit since the experimental setup does not enable continuous polarization of each individual cell.

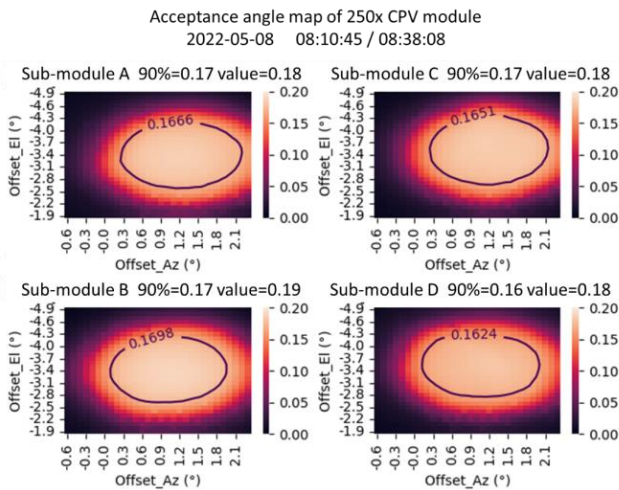

Figure 2: Acceptance maps of 250× CPV sub-modules obtained from data acquired on May 08th, 2022.

The SMU, the multiplexer and the tracker are driven using a Raspberry Pi with a dedicated controlling software written in Python.

A Razon+ weather station from Kipp'n Zonen installed ~200 m away from the test site is used to provide meteorological data (*i.e.* DNI and temperature). No measurement of the wind speed and direction is available at the test location. This parameter is therefore not considered here.

### D. Methodology

For all characterizations, we applied ASTM E2527 (American Standards of Technical Material) and the IEC 62670 (International Electrotechnical Commission), respectively, for data filtering and for characterization under standard conditions. The conditions on wind speed and spectrum are discarded. ASTM E2527 provides some restrictions needed to be applied on data for the regression such as: ambient temperature should be comprised between 10°C and 30°C, DNI should be more than 750W.m$^{-2}$, rejecting data if DNI deviation is more than 2% within 10 minutes [11]. As suggested by Muller *et al.* [11], we also reject data if diffuse radiation is above 140W.m$^{-2}$. IEC 62670-3 provides Concentrator Standard Operating Condition (CSOC) used to characterize CPV systems *i.e.* ambient temperature of 20°C, DNI of 900 W.m$^{-2}$ and Wind speed of 2 m.s$^{-1}$, AM1.5D spectrum [12]. The conditions on wind speed and spectral content are supposed to be verified (no measurement available on site) and no correction on cell temperature was applied for CSOC characterizations.

Two kinds of characterizations are performed here. First, we characterize the angle of acceptance of the sub-modules. For that, we map the short circuit current normalized to the DNI as a function of azimuth and elevation. This mapping is made by sweeping a frame of 3° x 3° in elevation and azimuth, respectively. A linear interpolation is used to fill missing points due to measurement variations. A contour corresponding to 90% of the maximum normalized short circuit current is then determined and plotted based on the interpolated normalized current map. Such acceptance maps are measured at various times of the day (from dawn to dusk). Using angular projection, the normalized short-circuit current is then plotted as a function of the angular deviation to the sun direction, resulting in a 1-dimension plot. The acceptance angle is finally determined by the average of the angles that yields 90% ±1% of the maximum normalized short circuit current.

TABLE I
ACCEPTANCE ANGLES OF 250× CPV SUB-MODULES OBTAINED FROM NORMALIZED SHORT CIRCUIT CURRENT AS A FUNCTION OF INCIDENCE ANGLE FROM DATA ACQUIRED ON MAY 08TH, 2022.

| Sub-module | Type | Acceptance angle (°) |
|---|---|---|
| A | Micro-beads | 0.79 ±0.02 |
| B | Micro-beads | 0.78 ±0.05 |
| C | Reference | 0.79 ±0.05 |
| D | Reference | 0.76 ±0.06 |

Second, we determine the I-V characteristics of each sub-module. The open-circuit voltage ($V_{OC}$) is measured first. Then, we sweep the cell voltage between 0V and $V_{OC}$. Eight points are linearly measured between 0V and 70% of $V_{OC}$, then 16 points are measured between 70% and 95% of $V_{OC}$, and 8 points between 95% of $V_{OC}$ and $V_{OC}$. The total sweep is then 32 values of cell current distributed on a voltage vector as previously described. Since it takes around 10 seconds per cell for tracker positioning and cell characterization, each cell is characterized every 3 minutes 16 seconds. From the I-V characteristics, we extract the short circuit current, the fill factor, and the maximum power point.

## III. RESULTS AND DISCUSSIONS

### A. Acceptance angle

The first step for the characterization of the sub-modules is the determination of their acceptance angle. Figure 2 presents the azimuth/elevation mapping of the short circuit current normalized to the DNI determined on May 8$^{th}$, 2022, between 8:10AM and 8:38AM. We can see that all sub-modules present a small offset of 1 to 1.5° in azimuth and 3 to 4° in elevation. This offset is a combination of the misalignment of the test module to the tracker and a misalignment of the tracker to the sun. Since the tracker is operating in MPPT mode for I-V characterization of each sub-module, this offset will automatically be corrected and has no significant impact on the cells characterization. Nonetheless, the offset is small enough for the MPPT mode to work starting from the astronomical position of the tracker. We can also see in Figure 2 that the 90% of the maximum short circuit current normalized to the DNI has an elliptical shape, characteristics of the symmetry of the system, which validates the cell to lens alignment procedure. Figure 3 shows the normalized short circuit current as a function of the angular deviation to the sun direction extracted from Figure 2 and from additional data collected all along the day of May 8$^{th}$, 2022. We can see that for the four sub-modules, more than 90% of the maximum short circuit current can be generated as long as the deviation to the sun direction is below 0.79° for sub-modules A and C and 0.78° and 0.76° for sub-modules B and D, respectively. A variation of ±0.02° to ±0.06° has been observed on these acceptance angles, due to the imperfect cylindrical symmetry of the sub-modules and tracker's resolution. Based on these measurements summarized in Table 1, we can say that all sub-modules have acceptance angle of ±0.78° and that there is no significant difference between the reference and the sub-modules with encapsulating ARC. This angle of acceptance exceeds the requirement of 0.4° for CPV systems [13]. Even though the angular acceptance at the cell level of the solar cells with micro-beads is improved compared to reference cells, this does not reflect into a larger acceptance angle at the module level.

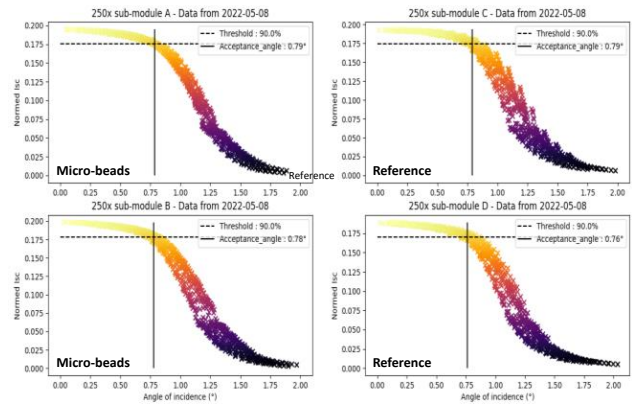

Figure 3: Normalized short circuit current as a function of 250× CPV sub-modules incidence angle obtained from data acquired on May 08$^{th}$, 2022.

TABLE 2
CONCENTRATION FACTOR AND CTM OF 250× CPV SUB-MODULES OBTAINED FROM THE CHARACTERIZATIONS CARRIED FROM MID-AUGUST TO MID-SEPTEMBER 2022.

| Sub-module | Type | $J_{SC}$ (mA/cm²) from EQE @Cell level @AM1.5D | $I_{SC}$ (mA) @Module level @900W.m$^{-2}$ | Effective concentration | Short circuit current CTM |
|---|---|---|---|---|---|
| A | Micro-beads | 13.24 | 196 | 226 | 90 |
| B | Micro-beads | 13.20 | 192 | 222 | 89 |
| C | Reference | 12.87 | 172 | 204 | 82 |
| D | Reference | 12.75 | 144 | 172 | 69 |

### B. Electrical characterization

The test module has been in operation in open loop tracking mode in the solar park of Université de Sherbrooke from Summer 2021 to Fall 2022. The MPPT mode has been implemented in mid-august 2022. Due to the data filtering imposed by ASTM E2527, only data acquired until mid-September 2022 can be used for determining the test module performance in CSOC, and the module has been dismounted on September 14$^{th}$, 2022.

Figure 4 presents the evolution of the short circuit current ($I_{SC}$) of the four sub-modules as a function of the DNI. As expected, we observe a linear variation of the short circuit current with the DNI for the four sub-modules. Based on the measured $I_{SC}$ and the current density ($J_{SC}$) derived from indoor EQE measurement before integrating the cells into the module, we can estimate the cell-to-module coefficient (CTM) on the short circuit current. For this, we calculate the ratio between the measured current at 900 W.m$^{-2}$ and the short circuit current calculated from EQE measurement, knowing the cell active area (0.0655 cm²). This ratio gives an effective sun concentration which should be as close as possible to the lens optical concentration ratio of 250×. The short-circuit current CTM ratio is then defined by the ratio of the effective concentration to 250×. Table 2 presents the current density derived from EQE, the current measured at the sub-module level for an irradiance of 900 W.m$^{-2}$, the effective concentration and the short circuit current CTM ratio for each sub-module. For all devices but device D, the CTM is between 82 and 90% which is close to record sub-modules presented in the literature [14]. Device D has a CTM of 69%, which will be discussed later. Due to this low performance, we use only device C as a reference in the following. The CTM losses are attributed to optical losses at the lens (transmission losses through the BK7 glass, reflection losses, comatic aberrations, etc...), optical losses by reflection of off-normal light on the cell, and to the larger cell temperature. The very high value we measure for the short circuit current CTM confirms that the cell and tracker alignment procedure enables a quasi-perfect cell/lens/sun alignment.

Sub-modules A and B with micro-beads encapsulating ARC present a short circuit current of 196 and 192 mA at 900 W.m$^{-2}$, 12 to 14% larger than the $I_{SC}$ measured on reference sub-modules C (169 mA). This relates also in a larger CTM (89 to 90% compared to 82% for the reference). This improvement exceeds the improvement of 2.6% measured in laboratory condition [7]. This can be explained by a wider angular distribution of the sunlight on the cell into a module compared to indoor conditions. Indeed, the gain of the micro-beads ARC compared to interferometric ARC increases when the angle of incidence deviates from the normal of the cell. In addition, solar spectrum received by the cell into a module may be different from AM1.5D, due to absorptions in the lens and due to atmospheric variations, which could enhance the sensitivity to the current of the top or middle-junction of the cells. The cell temperature in the module is also larger than in laboratory conditions which also contributes to increase the short circuit current when the top cell is limiting the current. Finally, we cannot rule out a faster degradation of the reference cells compared to the micro-beads ARC, which would lead to a higher short circuit current after one year of operation for the cells with micro-beads encapsulating ARC. However, preliminary data measured when the module was installed and the tracking mode was in open-loop showed also an improvement of the sub-module with micro-beads compared to the reference [8].

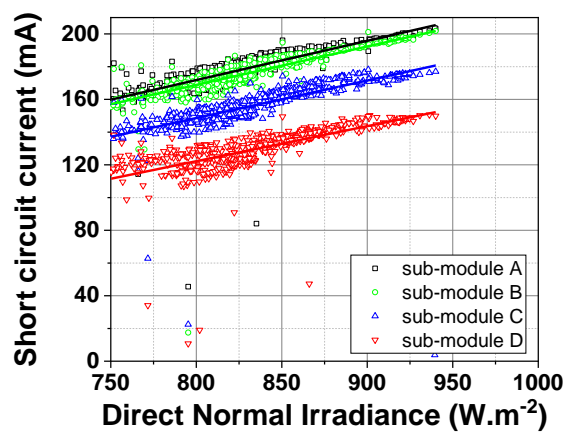

Figure 4: Short circuit current as a function of DNI of 250× CPV sub-modules after data filtering according to ASTM E2527 method described by the IEC 62670-3 standard (mid-August to mid-September 2022).

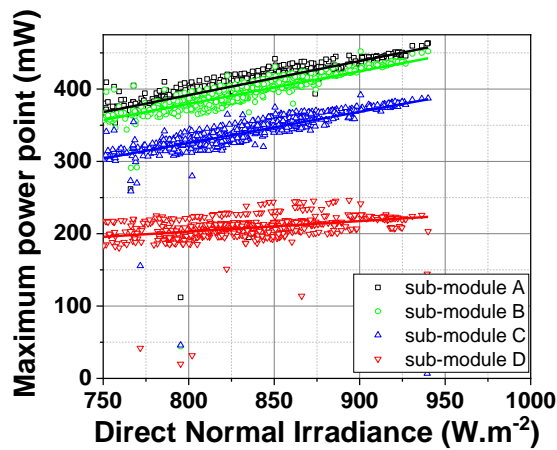

Figure 5: Maximum power point as a function of DNI of 250× CPV sub-modules after data filtering according to ASTM E2527 method described by the IEC 62670-3 standard (mid-August to mid-September 2022).

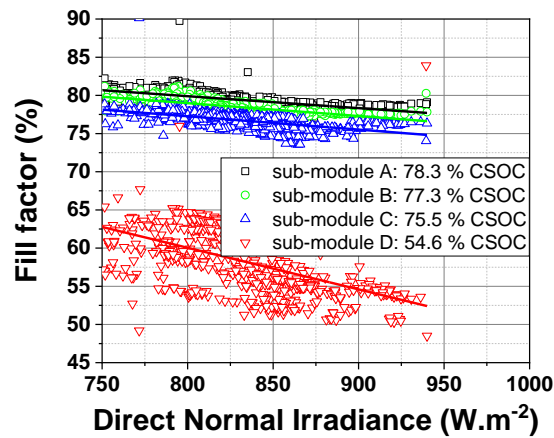

Figure 6: Fill factor as a function of DNI of 250× CPV sub-modules after data filtering according to ASTM E2527 method described by the IEC 62670-3 standard (August-September 2022).

Figure 5 presents the maximum power produced by the four sub-modules as a function of the DNI. As for the short circuit current, a linear variation of maximum power point (Pmax) as a function of DNI is observed. The maximum power point at CSOC is 439 and 425 mW corresponding to a sub-module efficiency of 29.7 and 28.8% for sub-modules A and B with micro-beads, respectively. This power is 15 to 19% larger than the maximum power measured on reference sub-modules C (369 mW, efficiency of 25%). The power and efficiency are reported Table 3.

The improvement for sub-module A and B performance compared to sub-module C is mostly due to the increase of the short circuit current, as explained in the previous part. An additional improvement is attributed to a larger fill factor (3.7% and 2.4% fill factor increase for sub-modules A and B, respectively) since the $V_{OC}$ is almost identical for all three sub-modules (2.85 V ±0.01V). The slightly lower $V_{OC}$ of sub-module C can be explained by a different assembly method between reference cells and cells with micro-beads encapsulating ARC, that may lead to different electrical and thermal resistances. It is therefore not attributed to the difference in anti-reflective coating.

The evolution of the fill factor with the DNI is presented Figure 6. We can see that the evolution of the fill factor with DNI is similar for sub-modules A, B and C, with a slope of -$1.7 \cdot 10^{-4}$ %/W.m$^2$ ±0.1%. On the contrary, sub-module D has a dramatically low fill factor (<55% at CSOC) with a slope of -5.4 %/W.m$^{-2}$. This low FF and higher degradation of the FF when the DNI increases are indicative of a defective device associated with a strong increase of the series resistance. The defectivity of the device is also confirmed by a degradation of the device performance (including its $I_{SC}$) between the preliminary measurements in open-loop mode (in 2021) compared to the measurements in MPPT mode (in August-September 2022). A visual inspection of cell D (not shown here) when the module was dismounted showed surface residues and degradation of the wire bonding pad that confirm the device defectivity.

## IV. CONCLUSIONS

We have integrated solar cells with a graded anti reflective coating (micro-beads semi buried into a PDMS layer) into a test module and rated the performance of the sub-modules under Concentrator Standard Operating Conditions (CSOC) after one year in the field. Their performance was compared with sub-modules made with reference cells with $AlO_x/TiO_x/SiO_2$ encapsulating ARC. Individual alignment of the solar cells and tracker positioning using maximum power point tracking enable a quasi-perfect cell / lens / sun

TABLE 3
MAXIMUM POWER POINT AND EFFICIENCIES OF 250× CPV SUB-MODULES OBTAINED FROM CHARACTERIZATIONS CARRIED OUT FROM MID-AUGUST TO MID-SEPTEMBER 2022.

| Sub-module | Type | Pmax (mW) @Module level @900W.m$^{-2}$ | Efficiency (%) |
|---|---|---|---|
| A | Micro-beads | 439 | 29.7 |
| B | Micro-beads | 425 | 28.8 |
| C | Reference | 369 | 25 |
| D | Reference | 217 | 14.7 |

alignment.

We demonstrated that even though the angular collection of light on the cell is improved by using a graded ARC, this does not reflect into a larger acceptance angle for the module. Electrical characterization of the sub-modules showed that the solar cells with the micro-beads ARC have 12 to 14% larger short circuit current than their $AlO_x/TiO_x/SiO_2$ encapsulating ARC counterpart, exceeding the expectations of 2.6% determined by indoor EQE measurement of the cells. We attributed this improvement to the broad angular distribution of light inside the module, compared to normal light incidence during EQE measurements, and to an actual spectrum received by the cell that differs from AM1.5D. A sub-module efficiency of 29.7 and 28.8% is reported for the sub-modules with micro-beads ARC. This is 3.8 to 4.7% absolute better than the reference sub-modules, mostly explained by a better light collection and to a lower extent by a different cell assembly method.

This paper demonstrates therefore that integrating a graded ARC, for instance using semi-buried silica micro-beads in PDMS, enables a significant efficiency gain for CPV modules compared to conventional multi-layer ARC. This improvement has been demonstrated after one year of operation in the field in harsh conditions in Quebec, promising therefore a high reliability compatible with long missions of CPV systems.

ACKNOWLEDGMENT

We acknowledge the support from STACE, MITACS and Quebec Ministère de l'Économie et de l'Innovation for their financial support. LN2 is a joint International Research Laboratory (IRL 3463) funded and co-operated in Canada by Université de Sherbrooke (UdeS) and in France by CNRS as well as ECL, INSA Lyon, and Université Grenoble Alpes (UGA). It is also supported by the Fonds de Recherche du Québec Nature et Technologie (FRQNT).

REFERENCES

[1] S. P. Philipps, A. W. Bett, K. Horowitz and S. Kurtz, *Current Status of Concentrator Photovoltaic (CPV) Technology*. United States: N. p., 2015. doi:10.2172/1351597.
[2] M. Alzahrani, A. Ahmed, K. Shanks, S. Sundaram, T. Mallick, *Solar Energy, 227 (2021) 321-333*.
[3] D.J. Aiken, Solar Energy Materials and Solar Cells, 64 (2000) 393-404.
[4] R. Homier, A. Jaouad, A. Turala, C.E. Valdivia, D. Masson, S.G. Wallace, S. Fafard, R. Ares, V. Aimez, *IEEE Journal of Photovoltaics, 2 (2012) 393-397*.
[5] P. García-Linares, C. Dominguez, O. Dellea, T. Kämpfe, Y. Jourlin, P. Besson, C. Weick, M. Baudrit, *AIP Conference Proceedings, 1679 (2015) 040004*.
[6] N. Shanmugam, R. Pugazhendhi, R. Madurai Elavarasan, P. Kasiviswanathan, N. Das, *Energies, 13 (2020) 2631*.
[7] G. P. Forcade, A. Ritou, P. St-Pierre, O. Dellea, M. Volatier, A. Jaouad, C.E. Valdivia, K. Hinzer, M. Darnon, *Progress in Photovoltaics: Research and Applications, 30 (2022) 132-140*.
[8] A. Ritou, P. St-Pierre, P.O. Provost, G. Forcade, C. Dubuc, O. Dellea, G. Hamon, M. Volatier, A. Jaouad, C.E. Valdivia, K. Hinzer, V. Aimez, M. Darnon, *AIP Conference Proceedings, 2550 (2022) 030004*.
[9] Thorlabs web site https://www.thorlabs.com/newgrouppage9.cfm?objectgroup_id=112&pn=LA1145
[10] Eko website https://www.eko-instruments.com/eu/categories/products/sun-trackers/str-22g-sun-tracker
[11] M. Muller, "Minimizing variation in outdoor CPV power ratings" : *2011 Reliability Workshop, Golden CO, USA.*
[12] "Concentrator Photovoltaic (CPV) solar cells and cell on carrier (CoC) assemblies – qualification" – International Standard IEC 62787, ed.1.0, 2021-02
[13] R. Mohedano, R. Leutz, "CPV Optics", *in: Handbook of Concentrator Photovoltaic Technology, 2016, pp. 187-238*.
[14] M. Steiner, A. Bösch, A. Dilger, F. Dimroth, T. Dörsam, M. Muller, T. Hornung, G. Siefer, M. Wiesenfarth, A.W. Bett, Progress in Photovoltaics: Research and Applications, 23 (2015) 1323-1329.